\documentclass[review]{elsarticle}
\journal{Elsevier}
\usepackage{lineno,hyperref}
\usepackage[utf8]{inputenc}
\usepackage[english]{babel}
\usepackage[T1]{fontenc}
\usepackage{amsmath}
\usepackage{amsfonts}
\usepackage{amssymb}
\usepackage{graphicx}
\usepackage{multirow}
\usepackage{float}
\usepackage{color}
\begin{document}
\begin{frontmatter}
\title{On the Mechanical Properties of Popgraphene-based Nanotubes: a Reactive Molecular Dynamics Study}
\author[UFPI]{W. H. S. Brand\~ao}
\author[UFPI]{A. L. Aguiar\corref{author}}
\ead{acrisiolins@ufpi.edu.br}
\cortext[author]{Corresponding author}
\author[UnB]{L. A. Ribeiro}
\author[UNICAMP1,UNICAMP2]{D. S. Galv\~ao}
\author[IFPI]{J. M. De Sousa}
\address[UFPI]{Department of Physics, Federal University of Piau\'i, Teresina, Piau\'i, Brazil.}
\address[UnB]{Institute of Physics, University of Bras\'ilia, Bras\'ilia, 70910-900, Brazil.}
\address[UNICAMP1]{Applied Physics Department, University of Campinas, Campinas, S\~ao Paulo, Brazil.}
\address[UNICAMP2]{Center for Computing in Engineering and Sciences, University of Campinas, Campinas, S\~ao Paulo, Brazil.}
\address[IFPI]{Instituto Federal do Piau\'i - IFPI, S\~ao Raimundo Nonato, Piau\'i 64770-000, Brazil.}

\begin{abstract}
Carbon-based tubular materials have sparked a great interest for future electronics and optoelectronics device applications. In this work, we computationally studied the mechanical properties of nanotubes generated from popgraphene (PopNTs). Popgraphene is a 2D carbon allotrope composed of $5-8-5$ rings. We carried out fully atomistic reactive (ReaxFF) molecular dynamics for PopNTs of different chiralities ($(n,0)$ and $(0,n)$) and/or diameters and at different temperatures (from 300 up to 1200K). Results showed that the tubes are thermally stable (at least up to 1200K). All tubes presented stress/strain curves with a quasi-linear behavior followed by an abrupt drop of stress values. Interestingly, armchair-like PopNTs ($(0,n)$) can stand a higher strain load before fracturing when contrasted to the zigzag-like ones ($(n,0)$). Moreover, it was obtained that the Young's modulus ($Y_{Mod}$) (750-900 GPa) and ultimate strength ($\sigma_{US}$) (120-150 GPa) values are similar to the ones reported for conventional armchair and zigzag carbon nanotubes. $Y_{Mod}$ values obtained for PopNTs are not significantly temperature dependent. While the $\sigma_{US}$ values for the $(0,n)$ showed a quasi-linear dependence with the temperature, the $(n,0)$ exhibited no clear trends.
\end{abstract}

\end{frontmatter}

\section{Introduction}

The experimental realization of graphene \cite{novoselov2004electric} created a revolution in materials science. In part due to this fact, there is a renewed interest in other possible 2D carbon allotropes, such as graphynes \cite{baughman1987structure, coluci2003families,solis2018structural}, which with graphene remain the only experimentally realized truly carbon-based 2D structures \cite{li2010architecture}. More recently, other theoretical structures were proposed: penta-graphene \cite{zhang2015penta}, phagraphene \cite{wang2015phagraphene}, twin-graphene \cite{jiang2017twin}, $\psi$-Graphene \cite{li2017psi}, popgraphene \cite{wang2018popgraphene}, among others \cite{doi:10.1002/pssb.201046583}. Popgraphene is a planar structure composed of $5-8-5$ carbon rings \cite{wang2018popgraphene}. DFT calculations indicated a metallic behavior, thermal and mechanical structural stability \cite{wang2018popgraphene}. The role of defects and thermal effects on the elastic properties of popgraphene membranes were recently investigated \cite{meng2019nanoscale, junior2020temperature} and evidenced a brittle behavior.

Similarly, as carbon nanotubes are generated rolling up graphene sheets, popgraphene nanotubes (PopNT) can be generated in the same way and are the subject of the present work. Indeed, many nanotubes based on $2D$ carbon structures were already proposed, such as graphynes \cite{coluci2003families}, penta-graphene \cite{chen2017mechanical,wang2017novel,quijano2017chiral,de2018mechanical} and phagraphene \cite{junior2020elastic}. To the best of our knowledge, a comprehensive study on the mechanical properties of PopNTs was not reported so far. 

In this work, we investigated the elastic properties and fracture dynamics of PopNTs with different chiralities (armchair and zigzag) and diameters. We carried out fully atomistic molecular dynamics (MD) simulations to address the mechanical behavior of these materials under uniaxial tensile loading at different temperatures (from 300 up 1200 K). 

\section{Methodology}
\label{sec2}

\begin{figure}[htb!]
	\centering
	\includegraphics[scale=0.7]{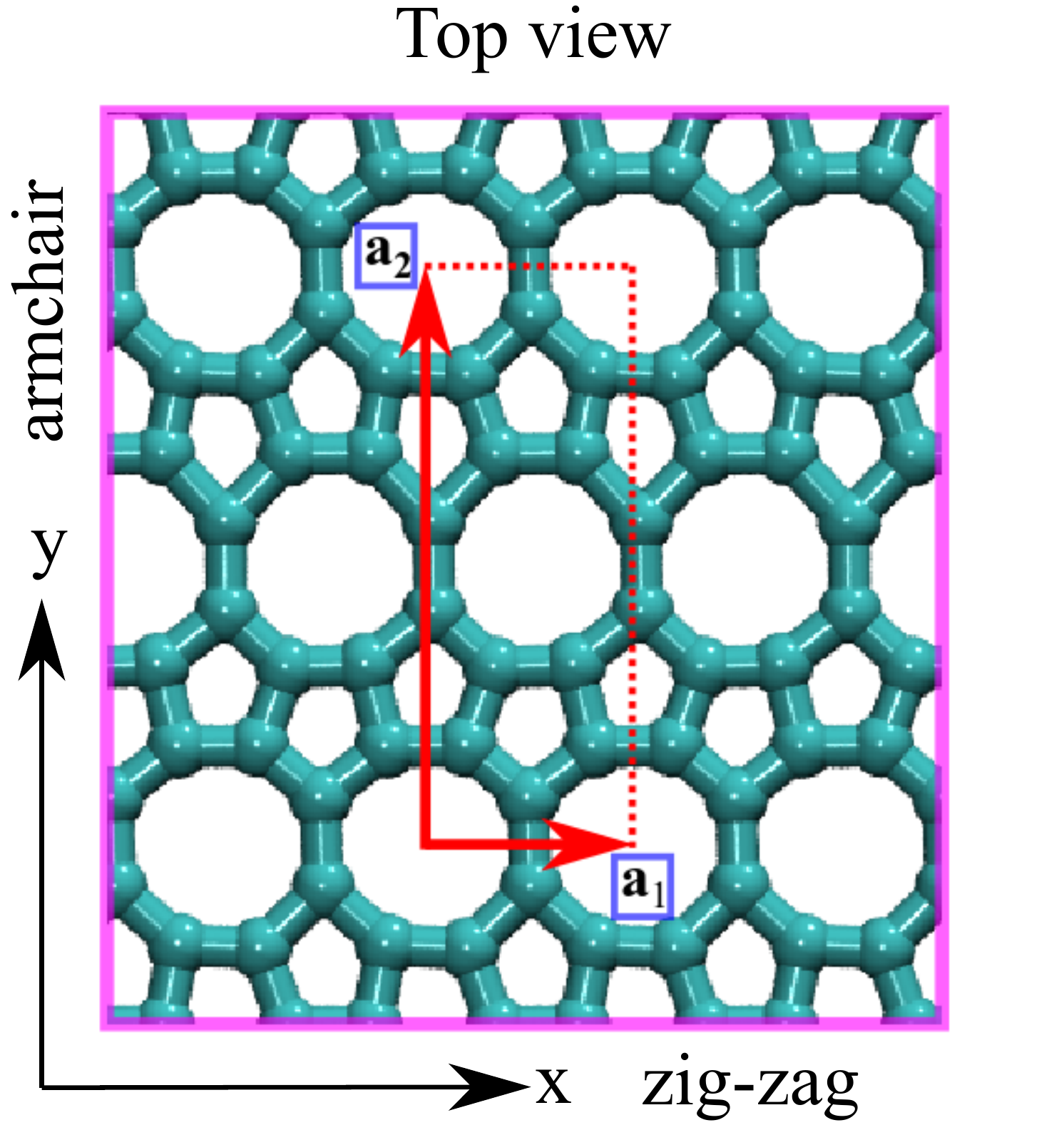}
	\caption{Schematic representation of a popgraphene monolayer. The rectangular geometry is defined by the vectors $\mathbf{a}_{1}=3.68$\r{A} $\mathbf{\hat{i}}$ and $\mathbf{a}_{2}=9.11$\r{A} $\mathbf{\hat{j}}$. The dashed rectangle represents the unit cell and the $x$ and $y-$axis are placed along with the zigzag and armchair directions, respectively.}
	\label{unit}
\end{figure}

The unit cell of a popgraphene membrane (which contains 12 carbon atoms) can be defined by a rectangle ($a_x=3.68$ \r{A} x $a_{y}=9.11$ \r{A}), as illustrated in Figure \ref{unit}. To obtain PopNTs, we followed the same procedure used to generate standard single-walled CNT (SWCNT) \cite{dresselhaus1995physics}. In this sense, the chiral vector ($\mathbf{C}_h$) is defined as
\begin{eqnarray}
\mathbf{C}_h=(n,m)=n\cdot\mathbf{a}_1+m\cdot\mathbf{a}_2,
\end{eqnarray}
where $\mathbf{a}_1=a_x\mathbf{\hat{i}}$ and $\mathbf{a}_2=a_y\mathbf{\hat{j}}$. A translational vector can be defined as the smallest vector orthogonal to $\mathbf{C}_h$ following $\mathbf{T}=(t_1,t_2)=t_1\cdot\mathbf{a}_1+t_2\cdot\mathbf{a}_2$, in which $t_1$ and $t_2$ are integers. 

\begin{figure}
	\centering
	\includegraphics[scale=0.26]{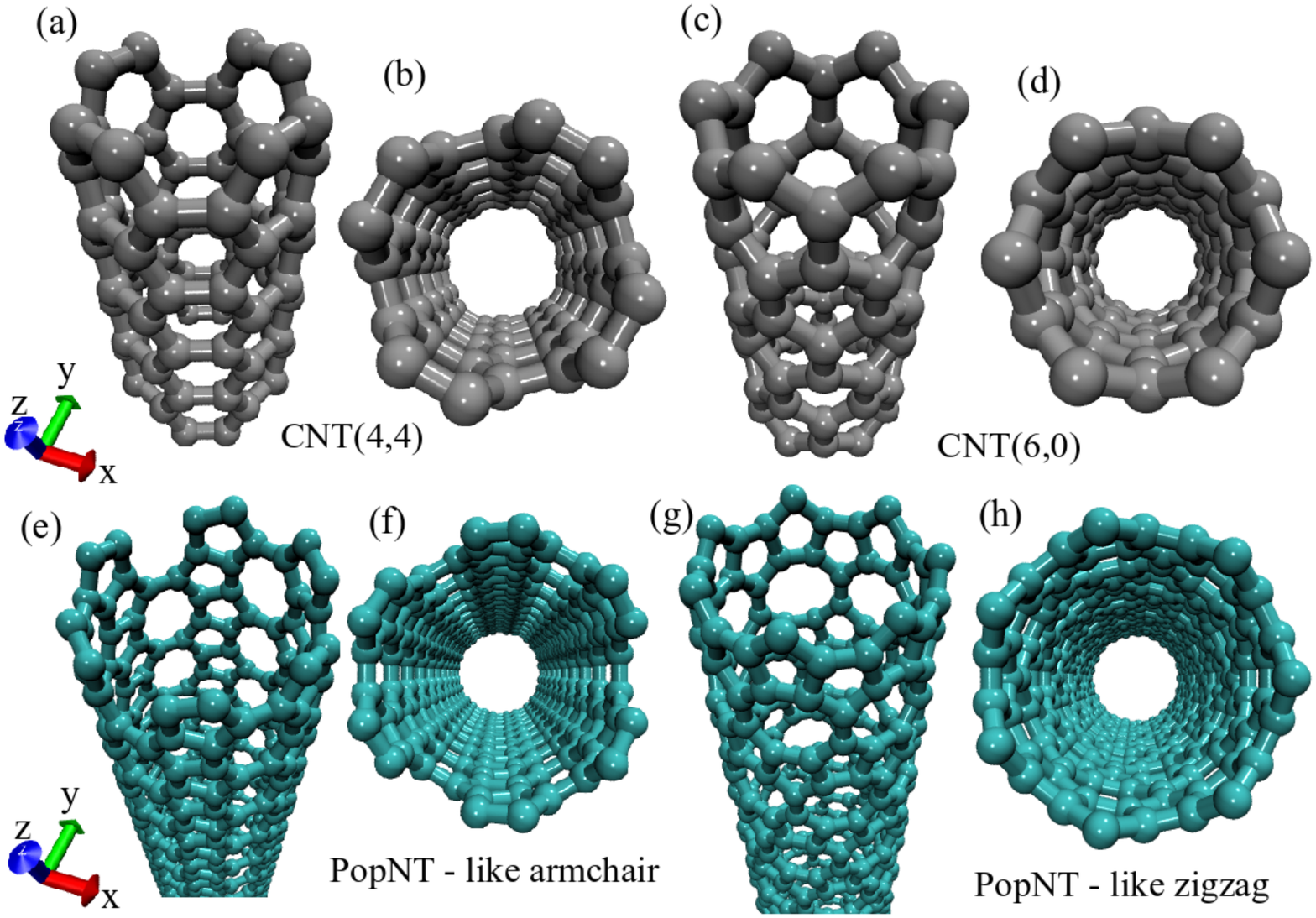}
	\caption{Atomic models of graphene (CNT) and popgraphene (PopNT) nanotubes. (a) and (b) shows an armchair $(4,4)$CNT. In (c) and (d) a zigzag $(6,0)$CNT. In (e) and (f) a armchair-like $0,3$PopNT, and in (g) and (h) a zigzag-like $(7,0)$PopNT.}
	\label{nanotubes}
\end{figure}

If we search for a condition that $\mathbf{C}_h \cdot \mathbf{T}=0$, we obtain $t_2/t_1 = -(na_x^2)/(ma_y^2)$. Due to the particular ratio $a_x/a_y \approx 0.4039$, no comensurability can be found for chiral $(n,m)$ combinations, and values of $t_1$ and $t_2$ can be very high for a given pair $(n,m)$ different from the $(n,0)$ and $(0,n)$ cases. To reduce the computational cost, the calculations were restricted to the $(n,0)$ and $(0,n)$ cases, named zigzag-like and armchair-like PopNTs, respectively, following the $x$ and $y$ directions in Fig. \ref{unit}. Similar to SWCNTs, achiral angle $\theta_c$ can also be defined as the angle between $\mathbf{C}_h$ and $\mathbf{a}_1$, which, in $(n,0)$ and $(0,n)$ cases, are restricted to $0^\circ$ and $90^\circ$, respectively, being associated to achiral tubes. These last achiral $(n,0)$ and $(0,n)$ tubes 
have translational vectors given respectively by $(0, 1)$ and $(1, 0)$. The number of atoms ($N_C$) in the nanotube unit cell is given by 12 times the number $N$ of $\mathbf{a}_1 \times \mathbf{a}_2$ rectangles within the area defined by  $\mathbf{C}_h$ and $\mathbf{T}$. Therefore, $N$ is obtained by dividing $\mathbf{C}_h \times \mathbf{T}$ per $\mathbf{a}_1 \times \mathbf{a}_2$, resulting only in
\begin{eqnarray}
 N_C^{(n,0)}=12\times\frac{|\mathbf{C}_h \times \mathbf{T}|}{|\mathbf{a}_1 \times \mathbf{a}_2|}=12n\\
 N_C^{(0,n)}=12\times\frac{|\mathbf{C}_h \times \mathbf{T}|}{|\mathbf{a}_1 \times \mathbf{a}_2|}=12n.
\end{eqnarray}
The length $L$ and radius $r$ of each generated PopNT as a function of $ \mathbf{T}$ and $\mathbf{C}_h$ can be calculated by
\begin{eqnarray}
L^{(n,0)}=|\mathbf{T}|=a_y\quad\textrm{and}\quad r^{(n,0)}=\frac{|\mathbf{C}_h|}{2\pi}=\frac{n}{2\pi}a_x\\
L^{(0,n)}=|\mathbf{T}|=a_x\quad\textrm{and}\quad r^{(0,n)}=\frac{|\mathbf{C}_h|}{2\pi}=\frac{n}{2\pi}a_y,
\end{eqnarray}
respectively. Analogous to the SWCNT case, the PopNT unit cell is obtained by rolling up the rectangle sector of the popgraphene membrane determined by $ \mathbf{T}$ and $\mathbf{C}_h$. Some representative PopNT and CNTs nanotubes (for comparison) can be visualized in Figure \ref{nanotubes}. 


Here, we carried out a systematic study on the mechanical properties of armchair- and zigzag-like PopNTs under tensile loading that was applied along the tube axial direction. To do so, we employed fully atomistic molecular dynamics (MD) simulations \cite{rapaport2004art,allen2017computer} by using the reactive interatomic ReaxFF potential \cite{mueller2010development, van2001reaxff,chenoweth2008reaxff,nielson2005development,rahaman2011development} as implemented in the LAMMPS code \cite{plimpton1995fast}. Importantly, the ReaxFF potential can describe breaking and bond formation and we have successfully used to study elastic properties and fracture dynamics of several nanostructures \cite{de2016mechanical,autreto2014site,de2016torsional,de2019elastic1,de2019elastic2}.  

In our simulations, we considered PopNTs with chiral index ranging within the intervals $4\leq n\leq 13$ and $2\leq n\leq 11$ for $(n,0)$ zigzag-like and $(0,n)$ armchair-like species, respectively. Periodic boundary conditions were assumed along the $z$ axis (see Fig. \ref{nanotubes}), replicating 4 and 10 unit cells for zigzag-like and armchair-like PopNTs, respectively. These replications yielded zigzag-like and armchair-like species with lengths of $36.44$ \r{A} and $36.80$ \r{A}, respectively, allowing us to contrast tubes of different chiralities but almost of the same length.

The nanotubes were thermalized using an NPT ensemble \cite{evans1983isothermal,salinas1997introduccao} --- with null pressure --- to eliminate residual tensions in the nanotubes before the beginning of the stretching loading. The temperature was kept constant (for each temperature studied) and controlled through an NVT ensemble by employing the Nos\'e-Hoover thermostat \cite{hoover1985canonical}. All MD runs were carried out using timesteps of $0.05fs$. The strain is applied by increasing the size of the simulation box along the periodic direction ($z$-axis), which is updated every $0.05fs$ with constant engineering tensile strain rate of the $10^{-6}/fs$. 

The Young's modulus is defined as $Y_{Mod}=\frac{d\sigma_{zz}}{d\epsilon_{z}}$, where $\sigma_{zz}$ is virial tensor stress and $\epsilon_{z}$ is the deformation along the axial direction. The stress tensor is defined as 
\begin{eqnarray}
\sigma_{ij} = \frac{\sum_{k}^{N}m_{k}v_{k_{i}}v_{k_{j}}}{\zeta} + \frac{\sum_{k}^{N}m_{k}r_{k_{i}}.f_{k_{j}}}{\zeta},
\label{Eq3}
\end{eqnarray}
with $\zeta = L\pi d_{tube}t$, considering a hollow cylinder as the volume of the PopNT, where $L$ is the tube length (\AA), $d_{tube}$ is the tube diameter (\AA), $t=3.35$ \r{A} is the standard thickness value used for carbon membranes \cite{wang2018popgraphene,pop2012thermal}, $N$ the number of neighborhood carbon atoms of atom with the $k$ index, $m_{k}$ is the mass of the carbon atom, $v$ is the velocity, $r$ the position of carbon atoms, and $f$ is the force per atom. 
\begin{eqnarray}
\sigma_{VonM}^{i} = \sqrt{\frac{(\sigma_{xx}^{i}-\sigma_{yy}^{i})^{2} + (\sigma_{yy}^{i}-\sigma_{zz}^{i})^{2} + (\sigma_{xx}^{i}-\sigma_{zz}^{i})^{2} + 6 \left[(\sigma_{xy}^{i})^{2} + (\sigma_{yz}^{i})^{2}  + (\sigma_{zy}^{i})^{2}\right]}{2}.} 
\label{Eq4}
\end{eqnarray}

The evolution of the average atomic stress distribution in the tubes under stretching is calculated using the von Mises stress tensor
\cite{fan2011multiscale,silvestre2015advanced,de2016mechanical,de2019elastic1} as showed in the Eq.\ref{Eq4}. Particularly, this stress tensor is used to qualitatively estimate how the stress accumulates/dissipates in the structures under strain. The $\sigma_{VonM}^{i}$ scalar values of the stress are computed for each atom during the stretching process.

\section{Results}
\label{sec4}

In Table \ref{tabMD} we present a summary of the structural and mechanical properties of the PopNT studied in this work. Young's modulus ($Y_{Mod}$, in units of GPa), critical strain ($\epsilon_{C}(\%)$), and the maximum stress $\sigma_{US}$ (which is called Ultimate Tensile Strength US(GPa)) values that PopNT can stand while being stretched before mechanical failure (fracture) are also listed.

\begin{figure}[htb!]
	\centering
	\includegraphics[scale=3.2]{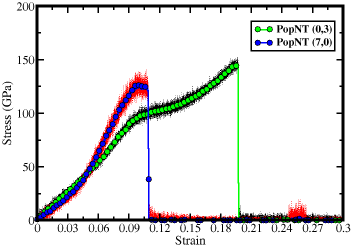}
	\caption{Stress-strain curves for armchair  $(0,3)$PopNT and zigzag $(7,0)$PopNT obtained at $300K$.}
	\label{stress}
\end{figure}

We estimated the Young's modulus values from the linear regime of the stress-strain curves \cite{turner2005xmgrace}. In Figure \ref{stress}, we present the results for $(0,3)$PopNT and $(7,0)$PopNT, which are representative cases of each chirality family.

One can see from stress-strain curves of $(0,3)$PopNT and $(7,0)$PopNT cases that, even with similar diameters, both tubes have different behaviors under uniaxial strain. $(7,0)$PopNT presents a hardening after $\epsilon_{C}\sim5\%$ up to $\epsilon_{C}\sim10\%$, where some $C-C$ bonds break (evidenced by the decrease in the stress values) until the complete failure, where an abrupt drop in the stress values takes place at the critical strain $\epsilon_{C}=10.84\%$. On the other hand, the elastic behavior of $(0,3)$PopNT presents two distinct regimes: a first quasi-linear regime up to $\epsilon_{C}\sim9\%$ and a second non-linear one after this value, which continues up to $\sim20\%$. After this critical value, the stress suddenly drops to zero, reaching the total fracture of the nanotube at $\epsilon_{C}=20.52\%$. 

The $Y_{Mod}$ (estimated up to $\epsilon_{C}\sim3\%$) and $\sigma_{US}$ (measured at critical strain) values for $(0,3)/(7,0)$ PopNTs are $845.12GPa/842.14GPa$ and $150.00GPa/129.42GPa$, respectively. Those $Y_{mod}$ and $\sigma_{US}$ values are similar to the ones reported for conventional armchair and zigzag CNTs using the same methodology\cite{de2019elastic1}. However, values for the critical strain obtained here of $20.52/10.84\%$ are significantly different from the corresponding CNT ones ($17.00/14.00\%$ for $(n,n)/(n,0)$ CNTs, respectively). Such differences for the critical strains can be explained by the different topologies of CNT and PopNT (see Figure \ref{nanotubes}). While conventional CNT are composed of densely packed carbon hexagons, PopNTs, in turn, present a porous morphology with $5-8-5$ carbon rings, which affects their elasticity/flexibility, as discussed below.

\begin{table}[htb!]
\caption{Structural PopNT parameters: tube radius ($r$) and number of atoms in supercell ($N_C$). The Young's modulus values $Y_{Mod}$ (in units of GPa) were estimated in the low-strain region $\epsilon_z=3\%$. The critical strain $(\epsilon_c)$ values were obtained from the corresponding ultimate tensile strength $(\sigma_{US})$ values (measured in GPa). The elastic properties were obtained at 300K for all PopNTs species studied here.}
\centering
\begin{tabular}{|c|c|c|c|c|c|c|}
\hline
\textbf{PopNT type}& \textbf{(n,m)} & r (\AA) & $N_C$ & \textbf{$Y^{300K}_{Mod}$} & $\boldsymbol{\epsilon_c}(\%)$ & 
$\boldsymbol{\sigma_{US}}$ \\ 
\hline
\multirow{10}{*}{zigzag $(n,0)$} 
& (4,0) &  $2.344$        &    $192$  &  748.06 & 10.27 & 126.63 \\ \cline{2-7}
& (5,0) &  $2.930$        &    $240$  &  778.19 & 10.61 & 132.21\\ \cline{2-7}
& (6,0) &  $3.516$        &    $288$  &  845.40 & 10.45 & 137.65\\ \cline{2-7}
& (7,0)  &  $4.102$       &    $336$  & 842.14 & 10.84 & 129.42\\ \cline{2-7}
& (8,0)  &  $4.688$       &    $384$  & 887.26 & 10.90 & 134.63\\ \cline{2-7}
& (9,0)  &  $5.274$       &    $432$  & 807.64 & 11.29 & 134.78\\ \cline{2-7}
& (10,0) &  $5.860$       &    $480$  & 834.72 & 10.94 & 129.72\\ \cline{2-7}
& (11,0) &  $6.446$       &    $528$  & 867.76 & 11.95 & 127.86\\ \cline{2-7}
& (12,0) &  $7.032$       &    $576$  & 850.88 & 12.41 & 127.88\\ \cline{2-7}
& (13,0) &  $7.618$       &    $624$  & 888.21 & 11.58 & 129.25\\ \hline
\multirow{10}{*}{armchair $(0,n)$} 
& (0,2)  &      $2.901$       &    $240$  & 839.10 & 21.15 & 154.32 \\ \cline{2-7}
& (0,3)  &      $4.352$       &    $360$  & 845.12 & 20.52 & 150.00\\ \cline{2-7}
& (0,4)  &      $5.803$       &    $480$  & 875.92 & 19.37 & 143.24\\ \cline{2-7}
& (0,5)  &      $7.253$       &    $600$  & 875.14 & 20.01 & 142.00\\ \cline{2-7}
& (0,6)  &      $8.704$       &    $720$  & 865.88 & 20.65 & 139.04\\ \cline{2-7}
& (0,7)  &      $10.154$      &    $840$  & 898.30 & 20.75 & 140.00\\ \cline{2-7}
& (0,8)  &      $11.605$      &    $960$  & 892.52 & 20.83 & 138.25\\ \cline{2-7}
& (0,9)  &      $13.056$      &    $1080$ & 896.55 & 20.94 & 136.57\\ \cline{2-7}
& (0,10) &      $14.506$      &    $1200$ & 908.66 & 20.89 & 136.46\\ \cline{2-7}
& (0,11) &      $15.957$      &    $1320$ & 899.89 & 20.64 & 136.14\\ \hline
\end{tabular}
\label{tabMD}
\end{table}
 
 
The anisotropy between $(0,3)$ and $(7,0)$ is present for the other tubes in Table \ref{tabMD}. While $Y_{Mod}$ values do not have a significant dependence on tube chirality and even no clear dependence on tube radius (which is proportional to $n$) is observed, both critical strain and ultimate stress values have significant dependence on the chirality. We observed that, in general, the $(n,0)$ PopNT have smaller critical strain and smaller $\sigma_{US}$ values than $(0,n)$ PopNT, which suggest $(0,n)$ PopNTs are more resilient materials. However, one should note that no evidence on the diameter dependence of critical strain values is observed. Interestingly, even if $\sigma_{US}$ values of $(n,0)$ PopNTs do not have significant dependence on diameter, $\sigma_{US}$ values found for $(0,n)$ PopNTs have a slight dependence on diameter, which means that the tensile stress needed for the fracture of small-diameter $(0,n)$ PopNTs is higher when compared to large diameter tubes. The origin of these differences can be addressed, by analyzing the geometric frames of the stretching loading process.

\begin{figure}[htb!]
	\centering
	\includegraphics[scale=30.0]{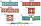}
	\caption{Representative MD snapshots for $(7,0)$ PopNT. (a) a side view of $(7,0)$PopNT at $\epsilon_z=0\%$; (b) a side view of $(7,0)$PopNT at $\epsilon_z=8.80\%$; (c) a zoomed view of (b); (d) $(7,0)$ PopNT at $\epsilon_z=10.87\%$ showing the beginning of the fracture; (e) a zoomed view of (d); (f) $(7,0)$PopNT fully fractured at $\epsilon_z=12.15\%$. The color bar indicates the von Mises stress values, where low (high) stress are represented in blue (red) color.}
	\label{snapshotsPopNTs70}
\end{figure}

\begin{figure}[htb!]
	\centering
	\includegraphics[scale=0.25]{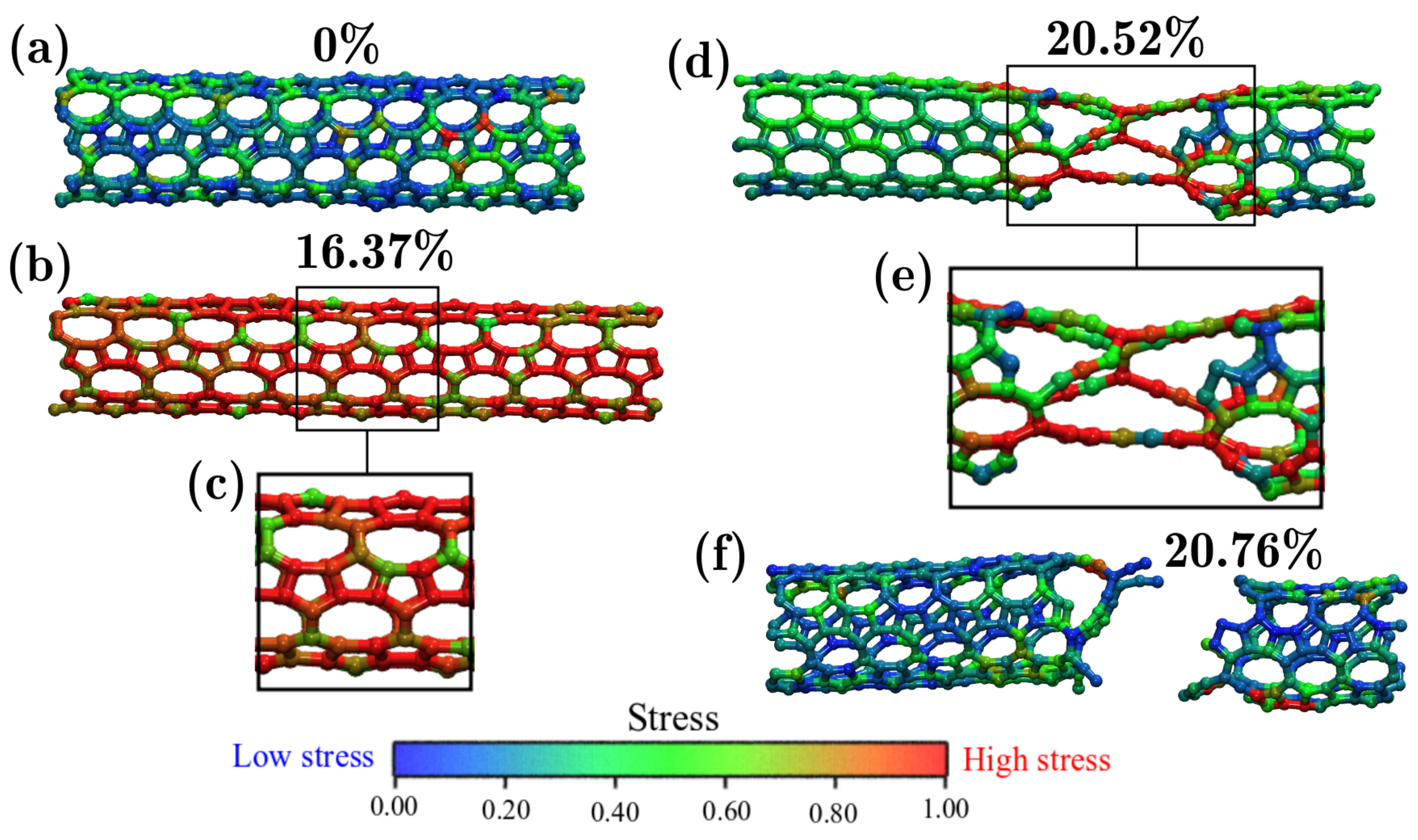}
	\caption{Representative MD snapshots for $(0,3)$ PopNT. (a) A side view of $(0,3)$PopNT at $\epsilon_z=0\%$. (b) A side view of $(7,0)$PopNT at $\epsilon_z=16.37\%$. (c) a zoomed view of (b);	(d) $(0,3)$PopNT at $\epsilon_z=20.52\%$ showing the beginning of the fracture. (e) a zoomed view of (d), and (f) $(0,3)$PopNT fully fractured at $\epsilon_z=20.76\%$. The color bar indicates the von Mises stress values, where low (high) stress are represented in blue (red) color.}
	\label{snapshotsPopNTs03}
\end{figure}

In Figure \ref{snapshotsPopNTs70} we present some representative MD snapshots and the corresponding von Mises stress values (Eq. \ref{Eq4}) for $(7,0)$PopNT. The stretching dynamics and the von Mises stress values were monitored until the complete tube failure, which is mechanically defined as the configuration when the nanotube is completely broken into two isolated pieces. Figure \ref{snapshotsPopNTs70}(a) presents the $(7,0)$PopNT equilibrated at room temperature with null uniaxial strain. The color differences in the von Mises values are due to random thermal fluctuations before starting the stretching process. Figure \ref{snapshotsPopNTs70}(b) illustrates a highly strained $(7,0)$ PopNT at $\epsilon_c=8.80\%$ and Figure \ref{snapshotsPopNTs70}(c) is a zoomed view of panel (b). Figure \ref{snapshotsPopNTs70}(d) illustrates the very first stage of the bond breaking that originates in the carbon ring composed of 8 carbon atoms. Figure \ref{snapshotsPopNTs70}(e) shows a zoomed view of panel (d). Finally, Figure \ref{snapshotsPopNTs70}(f) depicts the stage of the loading process in which the nanotube was completely fractured. In Figure \ref{snapshotsPopNTs03} we showed the corresponding cases for the $(0,3)$PopNT. 

As can be noted from Figures \ref{snapshotsPopNTs70} and \ref{snapshotsPopNTs03}, the fracture dynamics and the strain values for total fracture are different for $(7,0)$ and $(0,3)$ PopNT. Moreover, these figures also suggest that the $(0,3)$ PopNT takes longer to break with more pronounced formation of linear atomic chains in relation to the $(7,0)$ case.

\begin{figure}[htb!]
	\centering
	\includegraphics[scale=0.7]{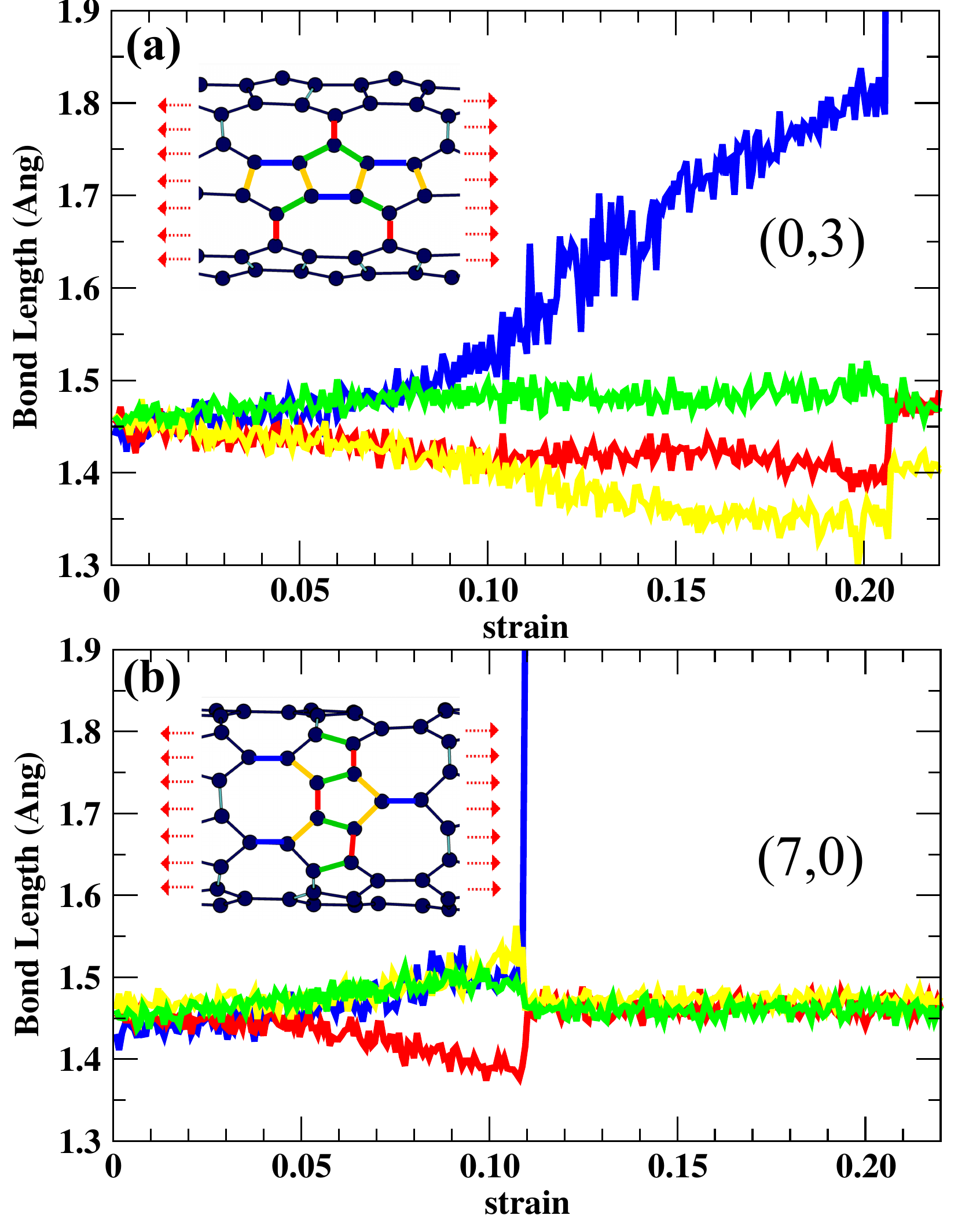}
	\caption{Bond length evolution as a function of the applied strain for and $(0,3)$ (a) and $(7,0)$ (b) PopNTs.}
	\label{strain}
\end{figure}

In order to explain these differences in the fracture dynamics between $(7,0)$ and $(0,3)$ cases, we present in Figure \ref{strain} a schematic model of the bond length evolution as a function of the strain values for some selected bonds. In Figure \ref{strain}(a), the inset illustrates the $(0,3)$ PopNT structure with a set of colored $C-C$ bonds. The blues ones are aligned with the stretching direction ($z$), in which the mechanical load is transferred to the $8$-carbon atoms ring. The green ones do not change significantly, while the yellow and red ones are compressed (thus partially compensating the blue stretching). This not happens for the $(7,0)$ PopNT (Fig.\ref{strain}(b)), wherewith the exception of the red ones, all bonds are stretched. For the $(0,3)$ PopNT this configuration allows the blue bonds to be longer stretched and explains why it breaks at larger strain values in comparison to the $(7,0)$ case.

\begin{table}[htb!]
\caption{Young Modulus values ($Y_{Mod}$) for all PopNTs studied for each temperature considered.}
\centering
\begin{tabular}{|p{25mm}|p{14mm}|p{14mm}|p{14mm}|p{14mm}|p{14mm}|}
\hline
\textbf{PopNT type} & \textbf{(n,m)} & \textbf{300 K} & \textbf{600 K} & \textbf{900 K} & \textbf{1200 K}  \\ 
\hline
\multirow{10}{*}{zigzag $(n,0)$} 
& (4,0) &  736.55 &	668.22 &	764.19 &	750.10 \\ \cline{2-6}
& (5,0) &  763.98 &	843.39 &	836.89 &	846.87\\ \cline{2-6}
& (6,0) &  836.74 &	942.44 &	848.75 &	718.48\\ \cline{2-6}
& (7,0)  &  842.14 &	838.63 &	884.52 &	887.11\\ \cline{2-6}
& (8,0)  &  890.66 &	860.99 &	833.69 &	856.77\\ \cline{2-6}
& (9,0)  &  807.64 &	821.84 &	937.40 &	813.17\\ \cline{2-6}
& (10,0) &  819.34 &	914.76 &	864.13 &	874.23\\ \cline{2-6}
& (11,0) &  857.25 &	973.12 &	963.58 &	895.44\\ \cline{2-6}
& (12,0) &  850.88 &	938.48 &	842.52 &	911.92\\ \cline{2-6}
& (13,0) &  888.21 &	856.02 &	 861.68 &	884.97\\ \hline
\multirow{10}{*}{armchair $(0,n)$} 
& (0,2)  &   837.50 &	793.69 &	943.00 &	967.72 \\ \cline{2-6}
& (0,3)  &   836.17 &	920.70  &	905.18 &	870.25 \\ \cline{2-6}
& (0,4)  &   878.44	&   843.75 &	922.48 &	1007.80 \\ \cline{2-6}
& (0,5)  &   869.32 &	877.03 &	893.82 &	993.20 \\ \cline{2-6}
& (0,6)  &   865.88	&   889.87 &	926.96 &	943.10 \\ \cline{2-6}
& (0,7)  &   892.95 &	900.00 &	913.72 &	920.59\\ \cline{2-6}
& (0,8)  &   892.88 &	898.55 &	934.29 &	922.49\\ \cline{2-6}
& (0,9)  &   900.83 &	900.21 &	958.97 &	930.05\\ \cline{2-6}
& (0,10) &   910.76 &	860.87 &	908.21 &	922.50\\ \cline{2-6}
& (0,11) &   899.31 &	911.54 &	922.89 &	921.11\\ \hline
\end{tabular}
\label{tabMD2}
\end{table}

\begin{figure}[htb!]
	\centering
	\includegraphics[scale=0.8]{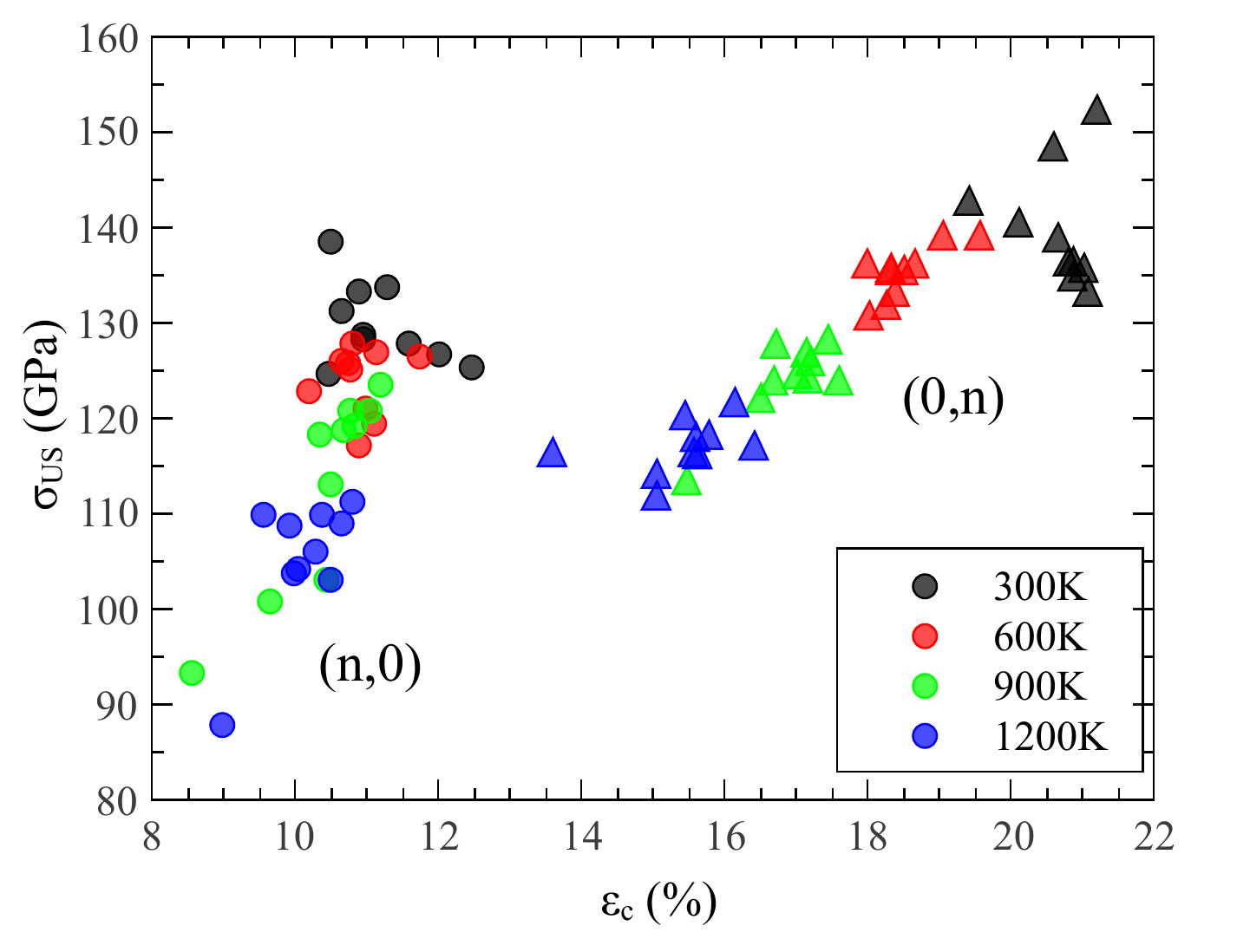}
	\caption{Ultimate stress values ($\sigma_{US}$) vs. critical strain ($\epsilon_c$) for several tube chiralities and temperatures. Circle symbols hold for zigzag-like PopNTs (n,0) while triangle symbols hold for armchair-like PopNTs (0,n).}
	\label{temperature2}
\end{figure}

We have also investigated the thermal effects on the elastic properties of PopNTs, considering the temperature range from 300 up to 1200K. The results are summarized in Table \ref{tabMD2} and Figure \ref{temperature2}. While $Y_{Mod}$ values are not significantly affected by temperature increasing (see Table \ref{tabMD2}) and no clear dependence on the tube diameter is observed. A significant reduction of ultimate tensile stress ($\sigma_{US}$) and the critical strain ($\epsilon_c$) was observed, especially for the more flexible $(0,n)$ PopNT (see Figure \ref{temperature2}). Such a reduction of $\sigma_{US}$ and critical strain values is expected due to
increasing of amplitude vibrations and also due to elongation of all $C-C$ bonds followed by an increase of $C-C$ bond-breaking due to the temperature. It can be seen from Figure \ref{temperature2} that while the $\sigma_{US}$ and $\epsilon_c$ values for $(0,n)$ tubes present an almost linear temperature dependence, no clear trend is observed for $(n,0)$ ones, which can be explained by the fact that $(0,n)$ tubes can stand larger strain values than the corresponding $(n,0)$ ones.

\newpage
\section{Conclusions}
\label{sec5}

In this work we present a comprehensive study on the mechanical properties of nanotubes generated from popgraphene (PopNTs), a carbon allotrope recently proposed. We carried out fully atomistic reactive (ReaxFF) molecular dynamics for PopNTs of different chiralities ($(0,n)$ and $(n,0)$) and/or diameters and at different temperatures (from 300 up to 1200K). Our results showed that the tubes are thermally stable (at least up to 1200K). All tubes presented stress/strain curves with a quasi-linear behavior followed by an abrupt drop of stress values, which is characteristic of a fast fracture. The $(0,3)$ tubes can stand a higher strain load before fracturing in relation to the $(n,0)$ ones, typically almost twice. With relation to the fracture dynamics, it was chirality dependent with the $(0,n)$ breaking later and with a more pronounced number of linear atomic chains.

The Young's modulus ($Y_{Mod}$) (750-900 GPa) and ultimate strength ($\sigma_{US}$) (120-150 GPa) values are similar to the ones reported for conventional armchair and zigzag carbon nanotubes (CNT) using the same methodology\cite{de2019elastic1} and they are not significantly dependent on the chirality and/or diameter. However, the values for the critical strain ($\epsilon_c$) for PopNTs are significantly chirality dependent, which is different from the corresponding CNT ones. With relation to the elastic behavior dependence with the temperature, the $Y_{Mod}$ values are not significantly dependent. While the $\sigma_{US}$/$\epsilon_c$ values for the $(0,n)$ show a quasi-linear dependence with the temperature, the $(n,0)$ exhibited no clear trends.

\section*{Acknowledgments}

This work was supported by the Brazilian Agencies CAPES, CNPq and FAPESP. J.M.S. and D.S.G. thank the Center for Computational Engineering and Sciences at Unicamp for financial support through the FAPESP/CEPID Grants $2013/08293-7$ and $2018/11352-7$. A. L. A. acknowledges CNPq (Process No. $427175/20160$) for financial support. W.H.S.B., A.L.A. and J.M.S thank the Laborat\'orio de Simulaç\~ao Computacional Caju\'ina (LSCC) at Universidade Federal do Piau\'i for computational support. L.A.R.J acknowledges the financial support from a Brazilian Research Council FAPDF and CNPq grants $00193.0000248/2019-32$ and $302236/2018-0$, respectively.

\bibliography{Pop.bib}
\end{document}